\def\ra{\rangle}
\def\la{\langle}
\def\t{\tilde}
\documentstyle[aps,prl,psfig]{revtex}
\begin{document}

\title{ Nonuniversal Critical Conductance 
Fluctuations of 
Chiral Surface States  
in the Bulk Integral Quantum Hall Effect
-- An Exact Calculation}
\author{ Yi-Kuo Yu }

\address{Physics Department, Case Western Reserve University
Cleveland, Ohio 44106-7079}
\date{\today}
\maketitle
\begin{abstract}
The chiral surface electrons in the bulk quantum Hall effect 
probably form the first extended system 
 in which conductance fluctuations can be calculated non-perturbatively
in the presence of disorder.
By use of the Kubo formula with appropriate 
boundary conditions, we calculate exactly the variance of
 conductance with non-perturbative methods. We find that the 
conductance fluctuations of this system are {\it nonuniversal}
and the variance of the conductance 
scales in a very peculiar way. This result can be checked
with exact computation using the Landauer-Buttiker formula and 
both methods show the same scaling behavior. 
We have also calculated the diffusion constant fluctuations exactly.
We find that the diffusion constant fluctuations vanish and thus
play no role in the conductance fluctuations.
\end{abstract}

Electronic systems under the influence of disorder
exhibit many intriguing phases most of which are still
not well understood\cite{Lee_P_85_a}. The understanding of the transition  
 of electronic states from localized\cite{Anderson_58} 
to extended is still far from complete. 
 In the standard 
scaling theory of localization\cite{Abrahams_79}, 
it is argued plausibly that the beta function 
$\beta(g)\equiv d\ln g/d\ln L$ is a function of the dimensionless
conductance $g$ alone. This one-parameter scaling
idea received further support  in the metallic regime from 
the experimental appearance of universal conductance fluctuations
(UCF)\cite{Umbach} in small {\it metals} as well as from
an interesting numerical simulation\cite{Stone_85} which revealed 
the UCF as 
a consequence of quantum interference. 
When the electronic states are localized, however,
the system generally has a very broad conductance distribution \cite
{Lee_P_85_a}. The average conductance thus might not seem enough
to fully characterize the system. Indeed, many-parameter scaling\cite{Shaprio}
has been proposed to take into account the broad conductance distribution
near the delocalization transition.
There is no method to determine from first principles 
 whether one-parameter or many-parameter scaling
is more appropriate except by calculating the higher cumulants
of the conductance, such as conductance fluctuations (CF), directly.
 So far CF calculations
have been done only perturbatively\cite{Lee_P_85_87,Altshuler}
 and are valid only in the metallic
regime. It is thus natural to ask will these
perturbative results hold to all orders of perturbation and 
what happens in the regime that is inaccessible to perturbative methods.
An exact conductance calculation in any physical system is thus 
important.  

We now introduce an interesting 2D anisotropic system,
 the chiral surface states of the bulk (multi-layered) 
integral quantum Hall effect(IQHE).
In the multi-layered quantum Hall sample,
 at the edge of each layer the electrons circulate in one 
direction only and can therefore be modeled as chiral Fermions\cite{Halperin}.
When the tunneling of these edge electrons in between the layers is allowed,
 the edge states then smear over  a two dimensional  sheet 
where electrons move 
 ballistically transverse to the field and diffusively
otherwise.
Bechgaard salts\cite{Bech}
and multi-layer hetrostructures\cite{hetro}
 are  candidates for this system.
The collection of these edge states turned out to be a novel metallic
phase and thus is also termed a chiral metal\cite{Fisher_a}.
This system is
 critical in the sense that the only length scale of the system
is the system size\cite{Chalker_b}.
One peculiar feature of the chiral metal 
is that it can retain conductance much smaller than 
$e^2/h$ while still scaling ohmically. This should be contrasted with
the ordinary metal for which ohmic scaling only happens at
large conductance. 

We shall focus on the limit
in which the length along
the chiral direction (circumference) $L_x$ is much greater than the
transverse dimension $L_z$; this corresponds to the
zero-dimensional regime defined by Balents {\it et al.}
\cite{Fisher_b}.  The finite $L_x$ case will be
discussed in a future publication\cite{YKY_b}.
Starting with linear response theory, we derive the Kubo formula suitable
for this system and calculate the disorder-averaged conductance as well as 
perform the {\it first exact 
conductance fluctuation calculation non-perturbatively}. We find that the 
CF are non-universal because the variance of the conductance is a function 
of the hopping amplitude $t$, which in turn is related to 
the system's diffusion constant. The CF 
scale with the system 
size as $L_x/L_z^2$. 
We also find that the disorder-averaged conductance scales ohmically: 
$\la g\ra \propto L_x/L_z$ where $\la \ra$ is used to 
denote the disorder average. Define $\delta g \equiv g-\la g\ra$
and $\Delta\equiv\la \delta g^2 \ra /\la g \ra^2$. We have 
$\Delta \sim 1/L_x \ll 1$ and thus $g$ is narrowly distributed.
Furthermore, $\Delta \to 0$ as $L_x \to \infty$ means that 
$g$ is self-averaging.
We have also calculated the average conductance using 
the Landauer-Buttiker(LB)
formula as well as the Einstein relation. These two results are 
identical. We have also calculated the diffusion constant
fluctuations {\it exactly} and find surprisingly that there 
are no diffusion constant fluctuations at all. This should 
be contrasted with the conventional result\cite{Altshuler}
for ordinary metals with dimensionality two
or lower
for which the major contribution
of CF is the diffusion fluctuation! This can 
be viewed as another salient feature 
of the novel chiral metallic phase. 
We can also calculate the CF by 
using LB formula, but the sums involved are difficult and 
are thus done numerically. The CF obtained from the LB
formula agree qualitatively with the Kubo calculation and show the same 
scaling behavior. One interesting feature worth mentioning is
that the CF vanish when $t=0$ or $t=2$ which
correspond to the situations of 
insulating and maximum-conducting respectively. We find
this happens both in LB type of calculation as 
well as in the Kubo type of calculation.

For simplicity, we consider the 
situation where only the first Landau level is filled, i.e. only one edge state
per quantum Hall layer.
The system thus 
consists of a collection of $1+1$ dimensional
ballistic states moving in the positive $x$ (chiral) direction,
 each described by a chiral Fermion, which are coupled
together by a nearest-neighbor hopping amplitude $t(n,x)$:
\begin{equation}
{\cal H}_0 = \sum_{n} \int dx \biggr[-\psi^\dagger_n\  i 
\partial_x^{\vphantom{\dagger}} \psi_n^{\vphantom{\dagger}} -
\bigg(t(n,x)\psi^\dagger_n \psi_{n+1}
^{\vphantom{\dagger}} + 
{\rm h.c.}\bigg)\biggr] . 
\end{equation}
There are two versions of the chiral model 
distinguished by the form of the hopping
amplitude $t(n,x)$. The first one we
shall call continuous hopping has $t(n,x)=t$ being a constant.
The second one which we shall call discrete hopping has $t(n,x)$ as
a sharply peaked function  at even(odd) integer $x$ for $n$
being even(odd)
and $t(n,x)$ being zero otherwise. 
 We shall focus only on the discrete hopping case.
The chiral fermion velocity $v$ is $1$ 
in our choice of units.  
The full Hamiltonian including random scattering is ${\cal H} =
{\cal H}_0 + {\cal H}_1$, where
\begin{equation}
{\cal H}_1 = \sum_n \int dx V_n(x) \psi^\dagger_n
\psi_n^{\vphantom{\dagger}} .
\end{equation}
 An illustration of this system is 
given in Fig.~1.
The random potential $V_n(x)$ is assumed to have 
zero mean. 
To make our discrete hopping model identical to the one 
used in ref.~\cite{Chalker_b},
 we require that the integral of the random potential over 
each link
$\int_{\rm link} V_n(x) dx \ {\rm mod}\ (2\pi)$ 
is a random variable distributed
uniformly over $(0,2\pi]$.  
In  the large $L_x$ limit,
due to the ballistic nature of the electron motion along the 
$x$ (chiral) direction
 the Green's function which is retarded in time must also be 
retarded in $x$\cite{Mathur}. The retarded Green's function 
with constant frequency $E$, $R_E$, satisfies the following equation 
\begin{eqnarray}
\bigg(-i\partial_x+V_{2n}(x)-E\bigg)R_E(2n,x;m,x')
+{\tilde \Lambda}_oR_E(2n-1,x;m,x')+{\tilde \Lambda}_eR_E(2n+1,x;m,x')
&=-\delta_{2n,m}\delta(x-x')\nonumber\\
\bigg(-i\partial_x+V_{2n-1}(x)-E\bigg)R_E(2n-1,x;m,x')
+{\tilde \Lambda}_oR_E(2n,x;m,x')+{\tilde \Lambda}_eR_E(2n-2,x;m,x')
 &=-\delta_{2n-1,m}\delta(x-x')\label{EQM}
\end{eqnarray}
where ${\tilde \Lambda}_{o(e)}(x)=t\sum_l^{\rm odd(even)}f(x-l)$ and 
$f(x)$ is a symmetric function sharply peaked around
zero with the condition that $\int f(x) dx=1$ \cite{footnote1}. 
 Note that the nonzero fixed frequency
$E$ can be gauged away by a simple local gauge 
transformation $G \to e^{iEx}G$. 

The retarded Green's function $R(n,x;n',x')$ obeying Eq.~(\ref{EQM}) can be 
expressed as a sum over paths of electron going 
from $(n',x')$ to $(n,x)$
\begin{equation}
R(n,x;n',x')=\sum_{p} \prod_{l\in p} \big[\chi_l(t) \exp (i\phi_l)\big].
\end{equation} 
As depicted in Fig.~1, the electron only
moves upwards along the links and can hop horizontally to its
neighboring layer across the nodes. Each path contributes an amplitude which
is a product of the phases $\phi=\int_{\rm link} V_n(x')dx'$ of the 
constituent links together with a prefactor $\chi(t)$ 
for each node which depends on whether the fermion hops to a neighboring
layer or remains in the same layer. The prefactors can be obtained
by integrating Eq.~(\ref{EQM}) across a node\cite{YKY_c}. An analogous 
expression can be written for the advanced Green's functions.

Let us first consider the disorder-averaged conductance 
which will illustrate the methods used here to calculate 
the CF. The disorder-averaged conductance has previously
been calculated for this model by Chalker and Dohmen\cite{Chalker_b}.
Using the standard linear response theory a Kubo formula 
can then be written that expresses the conductivity in terms of
the Green's function as
\begin{eqnarray}
&\sigma_{zz}(n,x;n',x')=-\Lambda_{n-1}(x)\Lambda_{n'-1}(x')
\bigg\{ \bigr[R(n-1,x;n'-1,x')A(n',x';n,x)
+R(n,x;n',x')A(n'-1,x';n-1,x) \nonumber\\ 
&\ \ \ \ \ \ \ -R(n,x;n'-1,x')A(n',x';n-1,x)
-R(n-1,x;n',x')A(n'-1,x';n,x)\bigr]+
\bigr[R\leftrightarrow A\bigr]\bigg\}e^2t^2/4\pi\label{COND}
\end{eqnarray}
where the $E=0$ subscript is suppressed from the Green's functions.
In deriving the above expression it is important to keep in
mind that the current operator along the non-chiral direction
is $\hat J_n(x)= iet \Lambda_{n-1}(x)[\psi^\dagger_n \psi_{n-1}-
\psi^\dagger_{n-1}\psi_n]$
where
$\Lambda_n(x)={\tilde \Lambda}_e(x)/t$ if $n$ is even and
$\Lambda_n(x)={\tilde \Lambda}_o(x)/t$ if $n$ is odd\cite{Fisher_a}.

Using the path integral expressions for the retarded and advanced 
Green's function, Eq.~(\ref{COND}) can be expressed as a double sum over
paths of retarded Green's function multiplied by the paths of advanced
Green's function.  Upon disorder-averaging, due to the random link phases,
only the diagonal terms, in which the retarded and advanced paths pair up,
 will survive in this double sum.
 Hence, the disorder-averaged conductivity
 is nonvanishing only when both $x$ and $x'$ are equal to
the same integer and $n=n'$\cite{YKY_c}
\begin{equation}
\la \sigma_{zz}(n,x;n',x')\ra=e^2t^2/2\pi\ \delta_{n,n'}\delta_{x,x'}.
\end{equation}
 This leads to
$\la g\ra=(L_x/2L_z)e^2t^2/2\pi$; the factor $L_x/2$ comes from 
the fact that $\Lambda_{e(o)}$ only sum over even(odd) integers along
the chiral direction. 

The disorder-averaged conductance can also be evaluated using 
the LB formula. This approach differs from the Kubo method in
that scattering boundary conditions at the probes are implemented exactly.
 To proceed we have to  calculate
the probability $|t_{ij}|^2$ of  finding outgoing current at $z=L_z, x=j$
while constant current is injected at $z=1,x=i$. The  conductance is given 
by $g = \sum_{i,j=1}^{L_x} |t_{i,j}|^2$. Note that in terms of diffuson
 $D(n,x;n',x')\equiv R(n,x;n',x')A(n',x';n,x)$, 
$|t_{i,j}|^2$ is given by $D(L_z,j;1,i)$. Calculating $\la g \ra$ is thus
equivalent to counting paths with appropriate weights. A direct
calculation\cite{YKY_c} thus shows that   
\begin{equation}\label{LBCOND}
\la g\ra ={e^2\over 2\pi}{t^2(L_x/2)\over t^2+(1-t^2/4)^2L_z}
\end{equation}
where the factor $L_x/2$ originates from the fact that along the chiral 
direction there is only one lead out for every two lattice spacings. 
The above result holds for all $t$ and agrees with the result
from the Kubo formula for small $t$. At $t=2$ where the system becomes
ballistic also in the non-chiral direction, the disorder-averaged conductance
saturates as one might expect from a Landauer conductance 
calculation. The result from Kubo formula, however, does not 
show this behavior as $t\to 2$.  It will be interesting to
understand the discrepancies between Kubo and Landauer results.
 For $t\neq 2$ and large $L_z$, 
$\la g \ra \simeq e^2t^2L_x/[4 \pi (1-t^2/4)^2 L_z]$ which reduces to
the result obtained in ref\cite{Chalker_b} with $t/(1+t^2/4)$ identified
as the $t$ used in ref\cite{Chalker_b}.

To calculate the average conductance in the third way, let 
us first consider the diffusion constant $D$ which is defined as the 
large $L$ limit of the quantity $\sum_n n^2 D(n,L)/2L$
where $D(n,x)\equiv D(n,x;n'=0,x'=0)$. 
We find that \cite{YKY_c}
 the disorder-averaged diffusion constant is 
given by $\la D \ra= t^2/2(1-t^2/4)^2 + {\cal O}(1/L)$. As $t\to 2$,
 it seems that $\la D \ra$ diverges. A careful analysis by having 
$L$ large but fixed shows that as $t\to 2$ one gets  $\la D \ra\sim L/2$.
This signifies that when $t\to 2$ the electron
motion becomes  ballistic in the non-chiral direction also. We 
also calculate exactly\cite{YKY_c}
 the diffusion constant fluctuations
and find them vanish for arbitrary $L$:
\begin{equation}\label{DFL}
\la (D-\la D\ra)^2\ra \equiv \sum_{n,m} n^2m^2[\la D(n,L)D(m,L)\ra
-\la D(n,L)\ra\la D(m,L)\ra]/4L^2=0.
\end{equation}
This indicates that the distribution of the diffusion constant is 
$\delta(D-D_0)$ independent of the disorder. If we wish to use the 
 Einstein relation to obtain conductivity in this case, 
 we only need to calculate the 
disorder-averaged density of states which 
is $1/2\pi$. Assuming Ohmic scaling, we then 
obtain the disorder-averaged conductance  $\la g\ra=e^2t^2L_x/
[4\pi L_z(1-t^2/4)^2]$ for $t \neq 2$. This result 
is exactly the same as the result from using the LB formula at large $L_z$
 and agrees with the result from the Kubo formula for small $t$.

The calculations of CF as well as the diffusion constant fluctuations
 involve disorder-averaging the product of 
four Green's functions. But as already
mentioned, since the Green's functions must pair up,
we can turn this average  
into a two-diffuson problem with contact interaction which comes
 from the possibility of 
exchanging paired-partners when two diffuson cross spatially.  
The quantity of central
importance is the probability of finding one diffuson at position
${\bf r}_1$ and the other at position ${\bf r}_2$
 for two diffusons starting at their
specified positions. In the 
large $L_x$ limit where the loops circled the circumference once or more 
can be
neglected, we can write down an evolution equation along the  chiral direction
for the interacting two-diffuson problem.
 We then perform a discrete Laplace transform and
introduce the center of mass momentum as well as the relative
momentum of the two diffusons.  The results are obtained
in Fourier space and are then converted back to real space\cite{YKY_c}.
A simpler version of this
calculation, although it appears in a different context, 
has also been done\cite{YKY_a}.

To calculate the CF exactly via the Kubo
 formula, we first observe that
 $g-\la g\ra$ is obtained by
 throwing away terms with $x=x'$ in Eq.~(\ref{COND}).
Note that the integral of $\Lambda_{n-1}(x)$ over $x$
gives rise to $\sum_l^{\rm even(odd)}$
for $n-1$ even(odd).
Furthermore, since
the contribution to the CF from $x>x'$ and
$x<x'$ will be the same due to the symmetry in Eq.~(\ref{COND}),
 we can work out the CF contributed
from $x>x'$ and then multiply it by two. Denote
$\t D(n,n')\equiv D(l_-,n;l'_+,n')$
 with both $l$ and $l'$ being
integers. After some calculations\cite{YKY_c},
in which the zero wave-vector mode 
along the non-chiral direction of diffuson was taken out, we obtain 
\begin{eqnarray}
\la \delta g^2\ra&=&(e^4t^4/4\pi L^4_z)\sum_{l=1}^{L_x}
\sum_{l'=0}^{l-1}\sum_{n(l),n'(l')}\big\la \t D(n-1;n'-1)
\t D(n;n')
+\t D(n-1;n')\t D(n;n'-1)\nonumber\\
&+& (1-t^2/4)^2t^2/(1+t^2/4)^4
\big[\t D(n-1;n'-1)
\t D(n;n'-1)+\t D(n;n')\t D(n-1;n')\big]
\big \ra\nonumber \\
&=&\big[e^4t^2(1-t^2/4)^2(1+\Gamma)L_x\big]/\big[4\pi^3(1+2\Gamma)L_z^2\big]\ \
\
{\rm when\ \ }L_x\gg L^2_z\gg 1
\label{CONDFL}
\end{eqnarray}
where $\Gamma=t^2(1-t^2/4)^2/(1+t^2/4)^4$.  For $t\neq 2$ and large $L_z$,
we find that $\la \delta g^2 \ra$ scales with the system size as
$L_x/L_z^2$. It is interesting to note that only in the $t\to 0$ limit
 where perturbative calculation being plausible, our
exact calculation agrees with the result of a perturbative calculation
\cite{Mathur}. For $t=2$ where electronic motion becomes ballistic 
also in the 
non-chiral direction, we find the CF vanish as physically expected.
 For general $t$, CF exhibit complicated $t$ dependence which is
not accessible to perturbative calculations.

Starting from LB formula $\la g^2 \ra$ contains 
disorder-averaged terms of products of 
two $|t_{i,j}|^2$. To show how this is done,
let us consider the case $i'>i$ and $j>j'$ for concreteness.
 The quantity $\la |t_{i,j}|^2
|t_{i',j'}|^2 \ra$ now can be written as 
\begin{equation}\label{LBCONDFL}
\la |t_{i,j}|^2|t_{i',j'}|^2 \ra =
\sum_{z_1,z_2=1}^{L_z}\la D(z_1,i';1,i)\ra 
 \la D(L_z,j;z_2;j')\ra \la D(z_2,j';z_1;i')
D(L_z,j';1,i')\ra.
\end{equation}
Following the methods outlined above, one can calculate
the disorder-averaged one-diffuson or two-diffuson explicitly.
  The sums over $z_1, z_2, i, i' j, j'$ turn out
to be quite complicated. We thus perform the sums numerically.
We have performed the calculations for $L_z$ up to
$10$ and $L_x$ up to $6000$. To extract the scaling behavior of the CF,
 we plot $(L_x/\la \delta g^2\ra)^{1/2}$ against $L_z$, and find that it
is very much a straight line for most values of $t$
 in good agreement with the results found in using the Kubo formula. 
The case for $t=0.31$ 
is shown in Fig.~2.  We have also check the specific cases where 
$t=0$ and $t=2$; and we see that the CF vanish in both cases in 
agreement with the Kubo calculation.

To conclude, we have performed the 
CF non-perturbatively by using
the path integral expression of the Green's function. 
We find that the CF are non-universal and scale in a peculiar way.
It has been speculated\cite{Stone_92}
 that UCF always appear for mesoscopic system with extended quantum states.
Our exact calculation shows that at least this system is an exception.

\acknowledgments
The author thanks T. P. Doerr for helpful discussions and
especially  H. Mathur for illuminating 
discussions, constant encouragement, and a critical reading of the manuscript.

\begin{figure}
\psfig{file=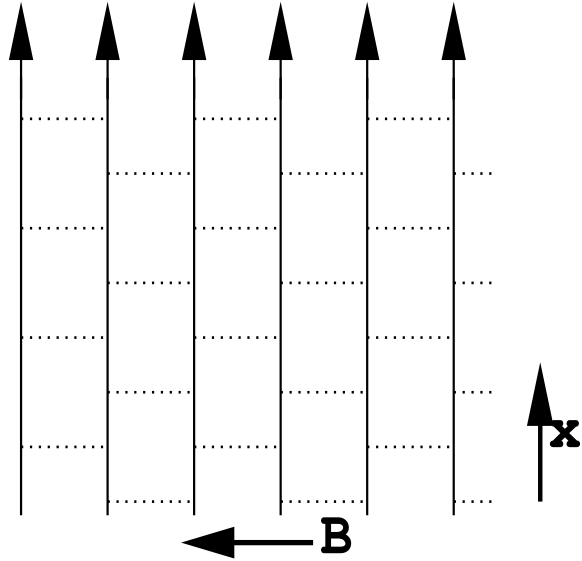,width=3.5in,height=3.5in}
\caption{The directed chiral network model.
 Full and dashed lines represent links and nodes respectively.}
\label{F1}
\end{figure}
\begin{figure}
\psfig{file=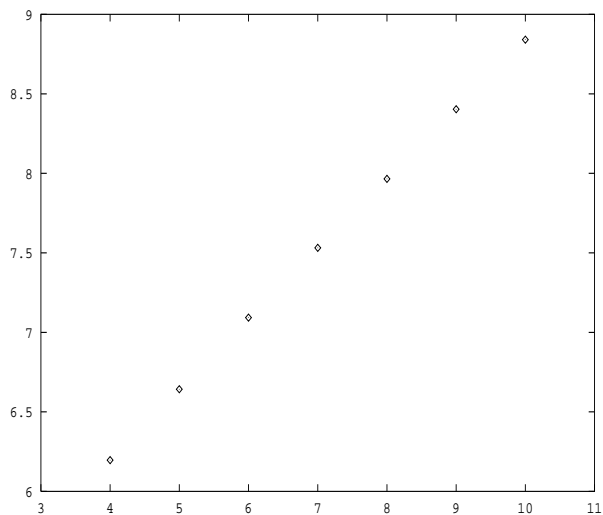,width=3.5in,height=3in,angle=-90}
\caption{The abscissa denotes $L_z$ and the ordinate marks the 
corresponding $(L_x/\la \delta g^2\ra)^{1/2}$. The hopping
parameter $t=0.31$.}
\label{F2}
\end{figure}


\begin{references}
\bibitem{footnote1} For an
arbitrary function $H(x)$ we define
 $\int f(x)H(x) dx = [H(0_-)+H(0_+)]/2$ to regularize the situation
where $H(x)$ is discontinuous at point $0$;
for $H(x)$ continuous at $0$ this definition is the same as
taking $f(x)\to \delta(x)$ at the end of calculation.
\bibitem{YKY_c} The detailed calculation will be described elsewhere.
\end{references}
\end{document}